\documentclass[10pt]{article}
\usepackage{geometry}
\usepackage{graphicx}
\usepackage{dcolumn}

\newcolumntype{P}[1]{>{\centering\arraybackslash}p{#1}}

\usepackage{float}
\usepackage[caption = false]{subfig}
\usepackage{siunitx}

\usepackage{setspace}

\emergencystretch=\maxdimen
\begin{document}

\begin{center}
\begin{large}
\textbf{Asymmetric modification of the magnetic proximity effect in Pt/Co/Pt trilayers\\by the insertion of a Ta buffer layer}
\end{large}
\end{center}

\begin{center}
Ankan Mukhopadhyay\textsuperscript{1\#}, Sarathlal Koyiloth Vayalil\textsuperscript{1\#}, Dominik Graulich\textsuperscript{2}, Imran Ahamed\textsuperscript{3},
\\Sonia Francoual\textsuperscript{4}, Arti Kashyap\textsuperscript{3}, Timo Kuschel\textsuperscript{2} and P S Anil Kumar\textsuperscript{1*}
\end{center}
\begin{center}
\begin{small}
\textit{\textsuperscript{1}Department of Physics, Indian Institute of Science, Bangalore 560012 India\\
\textsuperscript{2}Center for Spinelectronic Materials and Devices, Department of Physics,\\
Bielefeld University, Universitätsstraße 25, 33615 Bielefeld, Germany\\
\textsuperscript{3}School of Basic Sciences \& School of Computing and Electrical Engineering,\\
Indian Institute of Technology, Mandi 175005 India\\
\textsuperscript{4}Deutsches Elektronen-Synchrotron DESY, Notkestraße 85, 22607 Hamburg, Germany}
\end{small}
\end{center}
\renewcommand{\abstractname}{\vspace{-\baselineskip}}
\begin{abstract}

\noindent The magnetic proximity effect in top and bottom Pt layers induced by Co in Ta/Pt/Co/Pt multilayers has been studied by interface sensitive, element specific x-ray resonant magnetic reflectivity. The asymmetry ratio for circularly polarized x-rays of left and right helicity has been measured at the Pt $L_3$ absorption edge (11567 eV) with an in-plane magnetic field ($\pm158$ mT) to verify its magnetic origin. The proximity-induced magnetic moment in the bottom Pt layer decreases with the thickness of the Ta buffer layer. Grazing incidence x-ray diffraction has been carried out to show that the Ta buffer layer induces the growth of Pt(011) rather than Pt(111) which in turn reduces the induced moment. A detailed density functional theory study shows that an adjacent Co layer induces more magnetic moment in Pt(111) than in Pt(011). The manipulation of the magnetism in Pt by the insertion of a Ta buffer layer provides a new way of controlling the magnetic proximity effect which is of huge importance in spin-transport experiments across similar kind of interfaces.
\end{abstract}

\begin{center}
$^{*}$\textbf{anil@iisc.ac.in}\\
$^{\#}$equal contribution
\end{center}

\section*{\centering\normalsize{{I. INTRODUCTION}}}
\noindent Interfacial spin-orbit coupling in the ferromagnet-nonmagnet (FM/NM) systems promotes remarkable spin-related phenomena and interactions such as interfacial spin Hall effect \cite{r1,r2,r3,r5}, interfacial Dzyaloshinskii–Moriya interaction (iDMI) \cite{r8,r9,r10}, Rashba effect \cite{r11,r12} and spin–orbit torque (SOT) \cite{r13,r14}. All these effects simultaneously provide the electrical manipulation of the magnetization to control magnetization switching by current-driven domain wall motion. A static magnetic moment in the adjacent NM can be induced by the FM in certain FM/NM systems (e.g. Fe/Pd, Fe/Pt) \cite{r15,r16}. The phenomenon of a nominally paramagnetic material getting spin-polarized in presence of an adjacent FM or ferrimagnetic material by the exchange interaction, is known as magnetic proximity effect (MPE) \cite{r17,r18}. This effect can be commonly observed in materials such as Pt which is a Pauli paramagnet following the Stoner criterion \cite{r19}. \\
\noindent The MPE must be investigated to understand the magnetic properties of the atoms near the FM/NM interface. Earlier studies have shown that the structural properties of the atoms near the FM/NM interface determine the spin-transport properties \cite{r20,r21}. Furthermore, broken spatial inversion symmetry at the FM-NM interface can attribute to non-equivalent spin-transport properties (e.g. SOT, iDMI) \cite{r22,r23,r24,r25} between NM/FM (FM on top of NM) and FM/NM (NM on top of FM). Earlier studies show that in symmetric NM/FM/NM systems, SOT and iDMI can prevail due to different structural and/or magnetic properties at the NM/FM and FM/NM interfaces.\\
\noindent
MPE has been experimentally reported \cite{r26,r27,r28,r29,r30,r31,r32,r33,r34} mainly using x-ray magnetic circular dichroism (XMCD). In presence of FM metals such as Fe \cite{r29}, Ni \cite{r30,r31}, Co \cite{r32,r33,r34}, Pt has been reported to be spin-polarized and the absolute magnetic moment per Pt atom has been quantified using the XMCD sum rules in those Pt/FM bi- and multilayer systems. The XMCD technique becomes insensitive in extracting the absolute magnetic moment per atom of a specific element at buried interfaces for thicker layers in a multilayer thin film system \cite{r40}.Compared to XMCD, a new technique namely x-ray resonant magnetic reflectivity (XRMR) depends on the spin-dependent interference of x-rays reflected from the interfaces of a multilayer \cite{r41}. Depending on the energy of the x-rays used, the XRMR technique can detect the magnetic properties of multilayers with element- and depth-sensitivity. \\
\noindent
In this work, the induced spin polarization in top and bottom Pt layers due to Co layer in Ta/Pt/Co/Pt multilayers grown on Si/SiO\textsubscript{2} substrate has been investigated. We used the XRMR technique to quantify the magnetic moments in the two different Pt layers of the multilayer and found that the top Pt layer has larger induced magnetic moment than the bottom one that is adjacent to the Ta buffer layer. Furthermore, the induced magnetic moment in the bottom Pt layer reduces with increasing thickness of the Ta buffer layer. We have observed that the Ta buffer layer favours the growth of Pt(011) rather than Pt(111). Hence, by varying the thickness of the Ta buffer layer growth of the Pt layer can be modified, which in turn can control the MPE in the Pt layer due to the Co layer. Kuschel \textit{et al.} showed that the effective spin-polarized Pt thickness is independent of the thickness of the Pt layer by investigating FM/Pt bilayer systems \cite{r16}. In this work, we confirm this independency for Pt/FM/Pt multilayer systems. Experimental results are also compared with density functional theory calculations. In addition, the magnetic moment values observed by the XRMR technique is found to be comparable with the results obtained from a vibrating sample magnetometer (VSM). 
\section*{\centering\normalsize{{II. EXPERIMENTAL}}}
\noindent
Ta(\textit{t})/Pt(4)/Co(2)/Pt(2) multilayers (all the thicknesses are in nm) have been deposited by dc magnetron sputter deposition on Si substrates with 285 nm amorphous SiO\textsubscript{2} layer (3.8$\times$3.8 mm\textsuperscript{2}). The deposition has been done at room temperature (RT) with an Argon pressure of $5\times10^{-3}$ mbar in a high vacuum chamber with a base pressure below $5\times10^{-8}$ mbar. The deposition rates of Ta, Pt and Co were 0.022 nm/s, 0.047 nm/s and 0.023 nm/s, respectively. The thickness of the Ta buffer layer (\textit{t}) has been varied from 0 to 10.2 nm, where zero indicates the multilayer without any Ta buffer layer. During deposition the substrate has been rotated with 30 rotations/min in order to obtain uniform film thicknesses. VSM measurements at RT have been done on the multilayers with in-plane magnetic field up to $\pm1$ T.\\
\noindent
The XRMR and XMCD measurements were carried out at the resonant scattering and diffraction beamline P09 (EH1) of the third-generation synchrotron PETRA III at DESY (Hamburg, Germany) \cite{Sonia}. A six-circle diffractometer was used to carry out x-ray reflectivity (XRR) scans in $\theta$-$2\theta$ scattering geometry. XRMR is based on the magnetooptic change of optical properties of the investigated material. The complex refractive index of a material is given by $n=1-\delta+i\beta$, with dispersion coefficient $\delta$ and absorption coefficient $\beta$, connected via Kramers-Kronig relation. These coefficients vary by a fraction in magnetic materials depending on the relative orientation of the magnetization with respect to the incident beam and these variations are most pronounced at photon energies close to the absorption edge of the specific material. The theoretical background of XRMR has been described in Ref. \cite{r41} in detail.\\
\noindent
The fundamental aspect of XRMR is to determine the depth profile of the magnetooptic parameters $\Delta\delta$ and $\Delta\beta$ which correspond to the magnetic change of the dispersion $\delta$ and absorption $\beta$ coefficients, respectively. In our case, the XRR was measured at an off-resonant energy (11467 eV) and at the resonant energy (11567 eV) close to the whiteline of the Pt $L_3$ absorption edge in presence of an in-plane permanent magnet with a magnetic field of 158 mT at RT. The degree of circular polarization of the incident x-rays was $92\pm3$ \% as determined using an Au (111) analyzer crystal. A single 850-$\mu$m thick diamond plate at quarter wave plate condition was used to produce incident circularly polarized light. The reflected intensity of left ($I_L$) and right ($I_R$) helicity for each scattering vector \textbf{q} (XRMR) and energy (XMCD) was determined by switching fast the helicity of the incident circular polarization and averaging.\\
\noindent
Afterwards, the XRMR asymmetry ratio $\Delta$I$=$($I_L-I_R$)/($I_L+I_R$) was calculated from the collected data and fitted with an appropriate model. All the data processing and fitting have been done using the ReMagX analysis tool \cite{r43} following the recipe provided by Klewe \textit{et al.} \cite{r40}. The fitting algorithm for the nonmagnetic reflectivity data is based on the recursive Parratt formalism \cite{r44} with roughness described by the N\'{e}vot-Croce approximation \cite{r45}. The fitting algorithm for the asymmetry ratio is based on the Zak matrix formalism \cite{r46}. The tool models and varies the magnetooptic depth profiles, i.e., the in-depth distributions of $\Delta\delta$ and $\Delta\beta$, while keeping the absolute values for $\delta$ and $\beta$ unchanged. In the matrix formalism the roughness is also treated as an optical profile with a Gaussian-like distribution centred at the interface. The magnetooptic depth profile has been converted to Pt magnetic moment ($\mu$\textsubscript{B}/Pt) using the conversion factor obtained by \textit{ab initio} calculations \cite{r16}. \\
\noindent
The XMCD spectrum was collected using an energy dispersive silicon drift detector (VORTEX) synchronized with the piezo-actuators underneath the phase plates. Thus, allowing for the fluorescent photons for left and right circularly polarized incident beam to be counted separately at every point of the scan.\\
\noindent
The XRMR asymmetry ratio plotted against the scattering vector $q=\frac{4\pi\sin\theta}{\lambda}$ has been plotted in Fig. 1(a) for Ta(2)/Pt(4)/Co(2)/Pt(2). The profile changes the sign with the reversal of the magnetic field direction, which confirms its magnetic origin.

\section*{\centering\normalsize{{III. RESULTS AND DISCUSSION}}}
%

\begin{figure}[b]
\vspace{-4mm}
\centering
\includegraphics[scale=0.25]{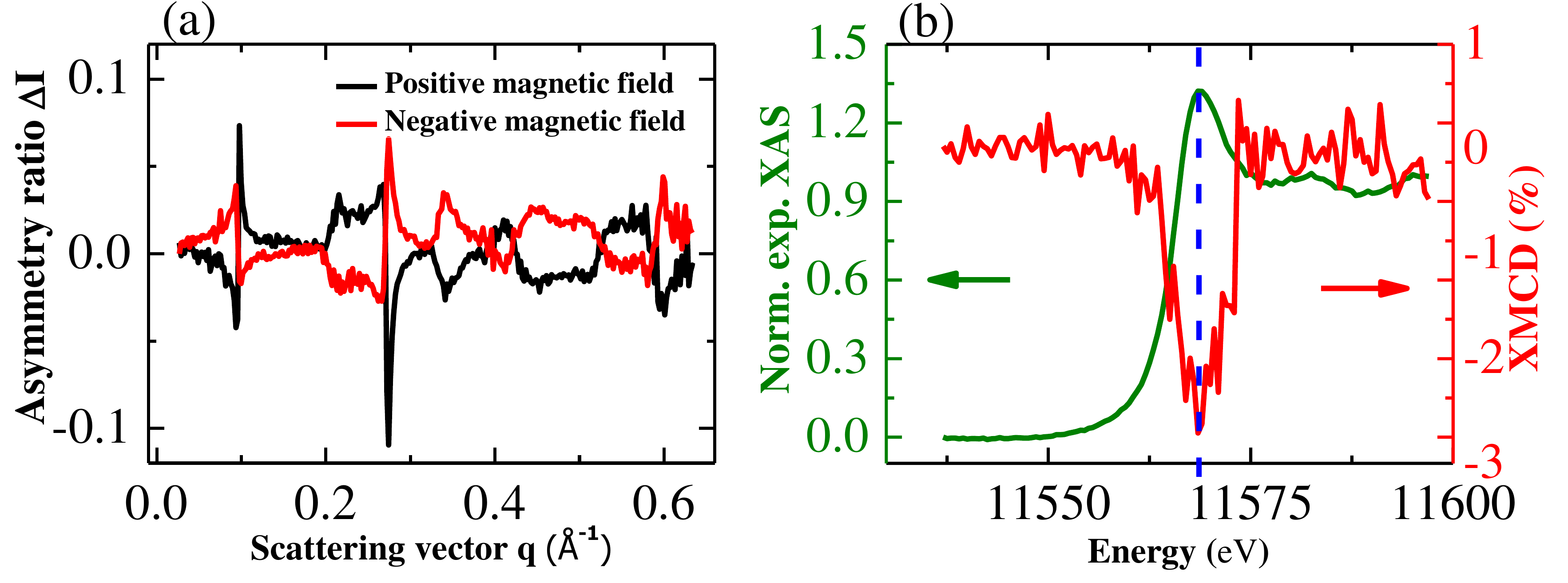}
\vspace{-4mm}
        \caption{(Color Online) (a) XRMR asymmetry ratio $\Delta$I(q) for bothmagnetic field directions at the resonant energy of 11567 eV. (b)
Experimental energy-dependent XAS spectrum (green curve)
and the XMCD signal (red curve). All data correspond to the
Ta(2nm)/Pt(4nm)/Co(2nm)/Pt(2nm) multilayer.}
    \label{fig:1}
\end{figure}

\begin{figure*}[t]
\centering
\hspace*{-2mm}
    \includegraphics[scale=0.35]{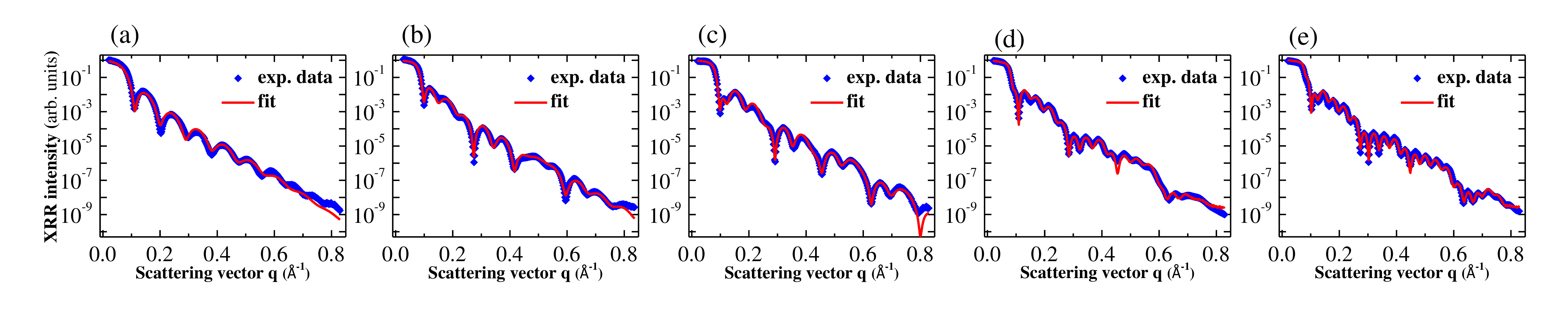}
        \vspace{-8mm}
    \caption{(Color Online)(a)-(e) The off-resonant (11467 eV) averaged XRR signals with fits for all the multilayers.}
    \label{fig:2}
\end{figure*}
\begin{table*}[htb!]
\centering
\begin{tabular}{|P{4.2cm}|P{0.8cm}|P{0.8cm}|P{0.8cm}|P{0.8cm}|P{0.7cm}|P{0.8cm}|P{0.7cm}|P{0.8cm}|P{0.7cm}|P{0.8cm}|} 
\hline
\textbf{ Multilayer}&\textbf{t}\newline \small SiO\textsubscript{2} & $\sigma$ \footnotesize bulk/Ta&\textbf{t\newline \small Ta}& $\sigma$ \small Ta/Pt&\textbf{t\newline \small Pt}& $\sigma$ \small Pt/Co&\textbf{t\newline \small Co}& $\sigma$ \small Co/Pt&\textbf{t\newline \small Pt}& $\sigma$ \small Co/Pt\\
\hline
\textbf{Pt(4)/Co(2)/Pt(2)}&bulk&\textbf{--}&0
\newline \tiny(without Ta)& 0.49
\newline \tiny(SiO\textsubscript{2}/Pt)&\textbf{3.40}&1.00&\textbf{1.63}&0.58&\textbf{1.60}&0.32\\
\hline
\textbf{Ta(2)/Pt(4)/Co(2)/Pt(2)}&bulk&0.47&\textbf{2.18}& 0.29&\textbf{3.65}&0.32&\textbf{2.16}&0.35&\textbf{1.59}&0.42\\
\hline
\textbf{Ta(4)/Pt(4)/Co(2)/Pt(2)}&bulk&0.46&\textbf{3.68}& 0.33&\textbf{3.69}&0.35&\textbf{1.66}&0.49&\textbf{1.81}&0.35\\
\hline
\textbf{Ta(7)/Pt(4)/Co(2)/Pt(2)}&bulk&0.49&\textbf{6.67}& 0.38&\textbf{3.71}&0.39&\textbf{1.57}&0.42&\textbf{1.92}&0.37\\
\hline
\textbf{Ta(10)/Pt(4)/Co(2)/Pt(2)}&bulk&0.42&\textbf{10.20}& 0.59&\textbf{3.37}&0.35&\textbf{1.57}&0.44&\textbf{1.90}&0.36\\
\hline
\textbf{Ta(3)/Pt(2)/Co(2)/Pt(2)}&bulk&0.52&\textbf{3.00}& 0.37&\textbf{2.04}&0.42&\textbf{1.97}&0.38&\textbf{1.77}&0.39\\
\hline
\end{tabular}
\vspace{-2mm}
\caption{Structural parameters, thickness t and roughness $\sigma$ expressed in nm of all the layers in each multilayer.}
\vspace{-5mm}
\label{tab:S1}
\end{table*}

\begin{figure*}[htb!]
\includegraphics[scale=0.43]{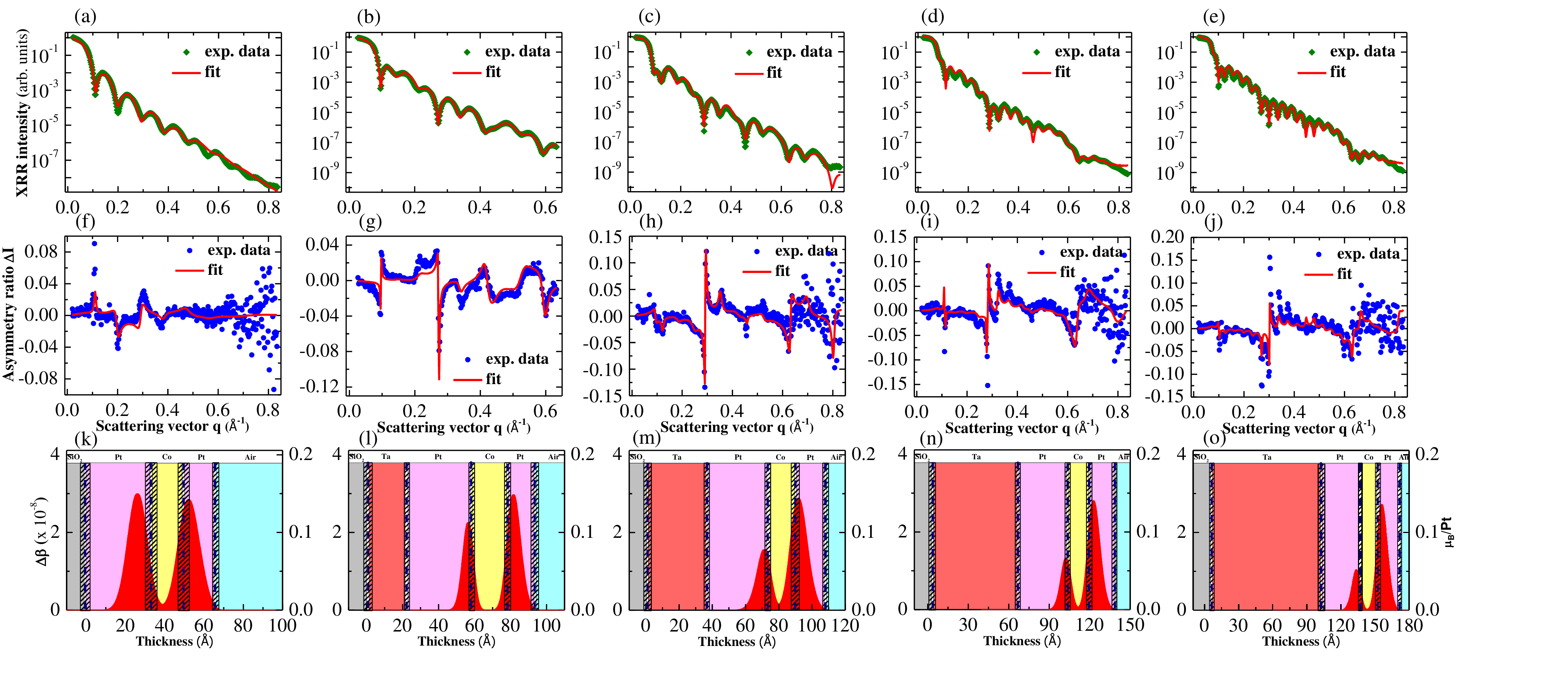}
\vspace{-10mm}
        \caption{(Color Online) (a)-(e) The resonant (11567 eV) averaged XRR signals with fits, (f)-(j) the corresponding measured and fitted XRMR
asymmetry ratios $\Delta$I(\text{q}) for the multilayers. (k)-(o) Magnetooptic depth profiles which were used to fit the XRMR asymmetry ratios. Red histograms
depict the in-depth distribution of the spin-polarized Pt atoms. Induced magnetic moment per Pt atom has been quantified from \textit{ab initio} calculation.
The dashed lines denote the corresponding interface position between the layers whereas the dashed rectangles between the layers denote the
corresponding interlayer roughness.}
    \label{fig:3}
\end{figure*}

\begin{table*}[hbt!]
\begin{tabular}{|P{5.8cm}|P{1cm}|P{1cm}|P{1cm}|P{1.3cm}|P{1.3cm}|P{1.2cm}|P{1cm}|} 
\hline
\textbf{ Multilayer}& \textbf{m\textsubscript{Co}} ($\mu$emu) &  \textbf{m$^{bottom}_{Pt}$} ($\mu$emu)& \textbf{m$^{top}_{Pt}$} ($\mu$emu)&  \textbf{m\textsubscript{XRMR}} ($\mu$emu)& \textbf{m\textsubscript{VSM}} ($\mu$emu)&\textbf{t$^{spin}_{bottom\hspace{1mm}Pt}$}\newline (nm)&\textbf{t$^{spin}_{top\hspace{1mm}Pt}$}\newline (nm)\\
\hline
\textbf{Pt(3.4)/Co(1.6)/Pt(1.6)}& 33.0& 16.1& 14.5& 63.6$\pm$3.7&67.0$\pm$5.5&1.30&1.24\\
\hline
\textbf{Ta(2.2)/Pt(3.7)/Co(2.2)/Pt(1.6)}& 43.7& 6.7& 11.7& 62.1$\pm$3.6& 64.5$\pm$6.0&0.94&0.98\\
\hline
\textbf{Ta(3.7)/Pt(3.7)/Co(1.7)/Pt(1.8)}& 33.6& 6.5& 13.7& 53.8$\pm$3.1& 59.5$\pm$4.3&1.10&1.17\\
\hline
\textbf{Ta(6.7)/Pt(3.7)/Co(1.6)/Pt(1.9)}&31.7& 4.5&13.1& 49.3$\pm$2.9&51.0$\pm$4.6&0.95&1.13\\
\hline
\textbf{Ta(10.2)/Pt(3.4)/Co(1.6)/Pt(1.9)}& 31.7&4.1& 13.1& 48.9$\pm$2.8&46.0$\pm$4.8&0.97&1.21\\
\hline

\end{tabular}
\begin{center}
\vspace{-5mm}
\caption{ \textbf{m\textsubscript{Co}}, \textbf{m$^{bottom}_{Pt}$} and \textbf{m$^{top}_{Pt}$} represents the magnetic moment
in the Co layer, the bottom and the top Pt layer respectively. \textbf{m\textsubscript{XRMR}}
and \textbf{m\textsubscript{VSM}} denotes the total magnetic moment of the multilayer
measured by XRMR and VSM, respectively.\textbf{t}\textsuperscript{spin} represents the spin-polarized thickness of the Pt layers.}
\end{center}
\label{table:2}
\vspace{-10mm}
\end{table*}

\begin{figure}[htb!]
\centering
        \includegraphics[width=0.45\textwidth] {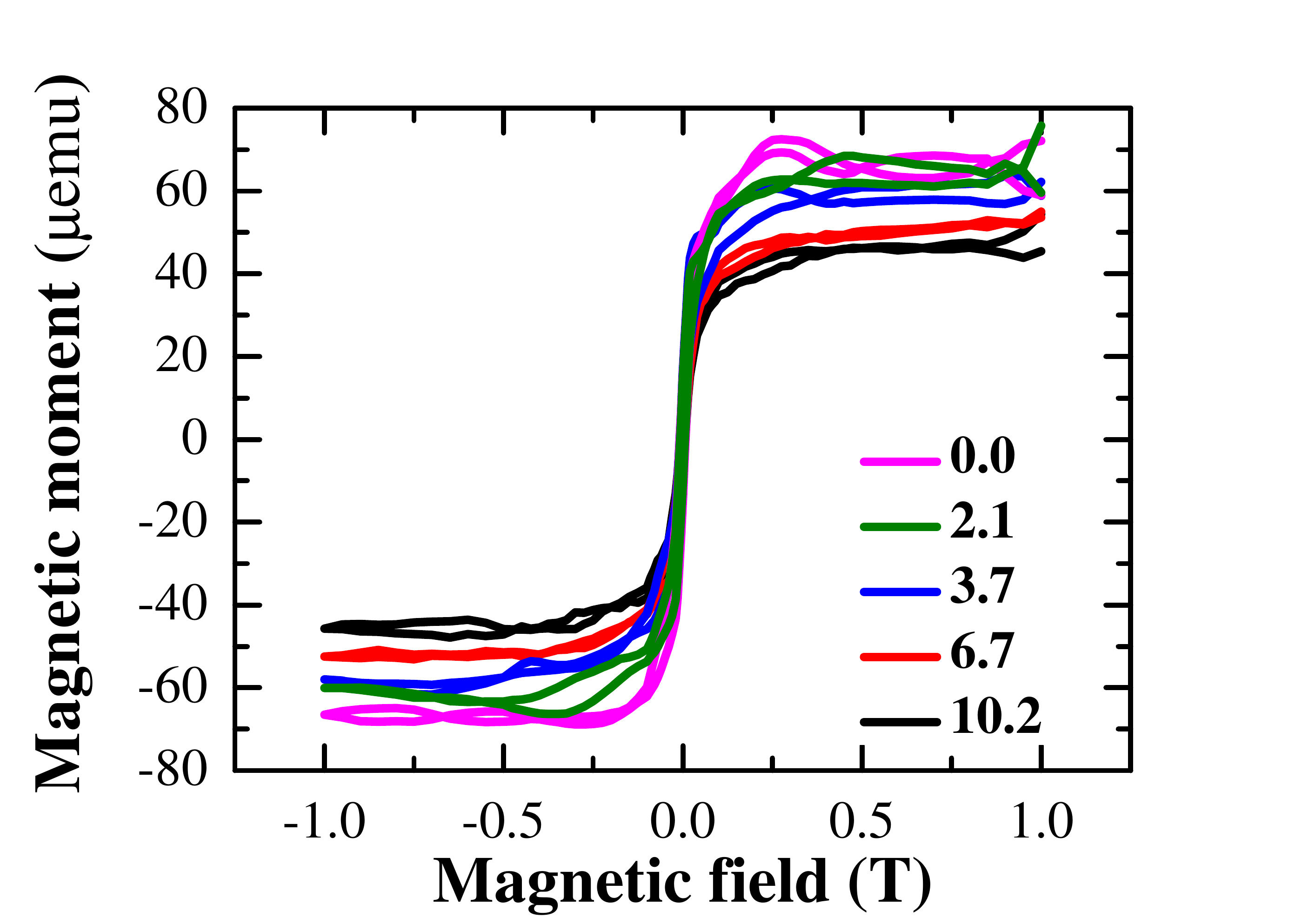}
        \vspace{-5mm}
        \caption{(Color online) Vibrating sample magnetometer measurement of the Ta(t)/Pt(4nm)/Co(2nm)/Pt(2nm) multilayers in in-plane geometry.}
        
        \label{fig:4} 
\end{figure}

\noindent
Experimental energy-dependent x-ray absorption spectrum (XAS) at the Pt $L_3$ edge normalized to the edge jump (green curve in Fig. 1(b) for Ta(2)/Pt(4)/Co(2)/Pt(2)) has been depicted along with the XMCD intensity (red curve in Fig. 1(b) for Ta(2)/Pt(4)/Co(2)/Pt(2)) to identify the energy with maximum dichroism. It can be confirmed that the maximum dichroism is close to the whiteline intensity of the $L_3$ absorption edge \cite{r27} comparable to recent \textit{ab initio} calculations \cite{r16} and experimental studies using fixed-\textit{q} scans \cite{r47} and XMCD \cite{r48}. We have chosen the resonant energy 11567 eV for collecting the XRMR data. The whiteline intensity (ratio of absorption maximum and edge jump) is 1.32 which indicates the metallic states of Pt (1.25 for metallic Pt, 1.50 for PtO$_{1.36}$ and 2.20 for PtO$_{1.6}$) \cite{r49}. \\
\noindent
The footprint-corrected normalized XRR signal of the multilayer collected at an off-resonant photon energy of 11467 eV has been plotted against the scattering vector \textit{q} (see Figs. 2(a)-(e)). The structural parameters have been obtained by fitting the off-resonant XRR data of each multilayer using the Parratt formalism and are tabulated in Tab. I. After determining the layer thicknesses and interlayer roughnesses in a certain multilayer by the off-resonant fit (11467 eV), the averaged resonant (11567 eV) XRR curve has been fitted (indicated in Figs. 3(a)-(e)) with the fixed structural parameters to obtain the optical parameters $\delta$ and $\beta$ of all the layers. The structural parameters obtained from the off-resonant XRR fit and the optical parameters obtained from the resonant XRR fit have been used to fit the XRMR asymmetry ratio by varying only the $\Delta\beta$ depth profile, keeping $\Delta\delta$ to be zero. In earlier studies \cite{r16,r40}, it has been shown theoretically by Kuschel \textit{et al.} and experimentally by Klewe \textit{et al.} that for the Pt absorption edge $\Delta\delta$ becomes zero at the maximum of $\Delta\beta$. The measured XRMR asymmetry ratios for the multilayers are shown in Figs. 3(f)-(j) accompanied by their respective fittings. The magnetooptic depth profiles of $\Delta\beta$ are shown in Figs. 3(k)-(o). These profiles were generated by a Gaussian function at each of the Pt/Co interface, convoluted with the roughness profiles of the respective interface \cite{r40}.

\noindent
The full width at half maximum (FWHM) of these Gaussian-like profiles has been calculated to quantify the effective thickness of spin-polarized Pt layers at the two interfaces. By comparing the experimental fit values of $\Delta\beta$ with the \textit{ab initio} calculations \cite{r16}, the magnetic moment per spin-polarized Pt atom at the maximum of these profiles has been determined. The magnetic moment contribution m\textsubscript{XRMR} from the Co layer with a saturation magnetization of 1400 emu/cm\textsuperscript{3} at RT \cite{r51} and the two Pt layers in a certain multilayer has been calculated by determining the area under the curve of the magnetooptic profiles. The saturation magnetic moments m\textsubscript{VSM} for all the multilayers have been extracted from their respective VSM measurements at RT (see Fig. 4). All the relevant extracted data have been summarized in Tab. I. It can be readily seen from Tab. I that m\textsubscript{XRMR} and m\textsubscript{VSM} are very much comparable. Hence, the total magnetic moment of a NM/FM/NM system can be considered as summation of the magnetic moment of the FM layer and the induced magnetic moment in the NM layer by the FM layer. Hence, the enhancement in magnetic moment of a multilayer can be attributed to the MPE.\\

\begin{figure}[t]
\vspace{-5mm}
\centering
\includegraphics[scale=0.35]{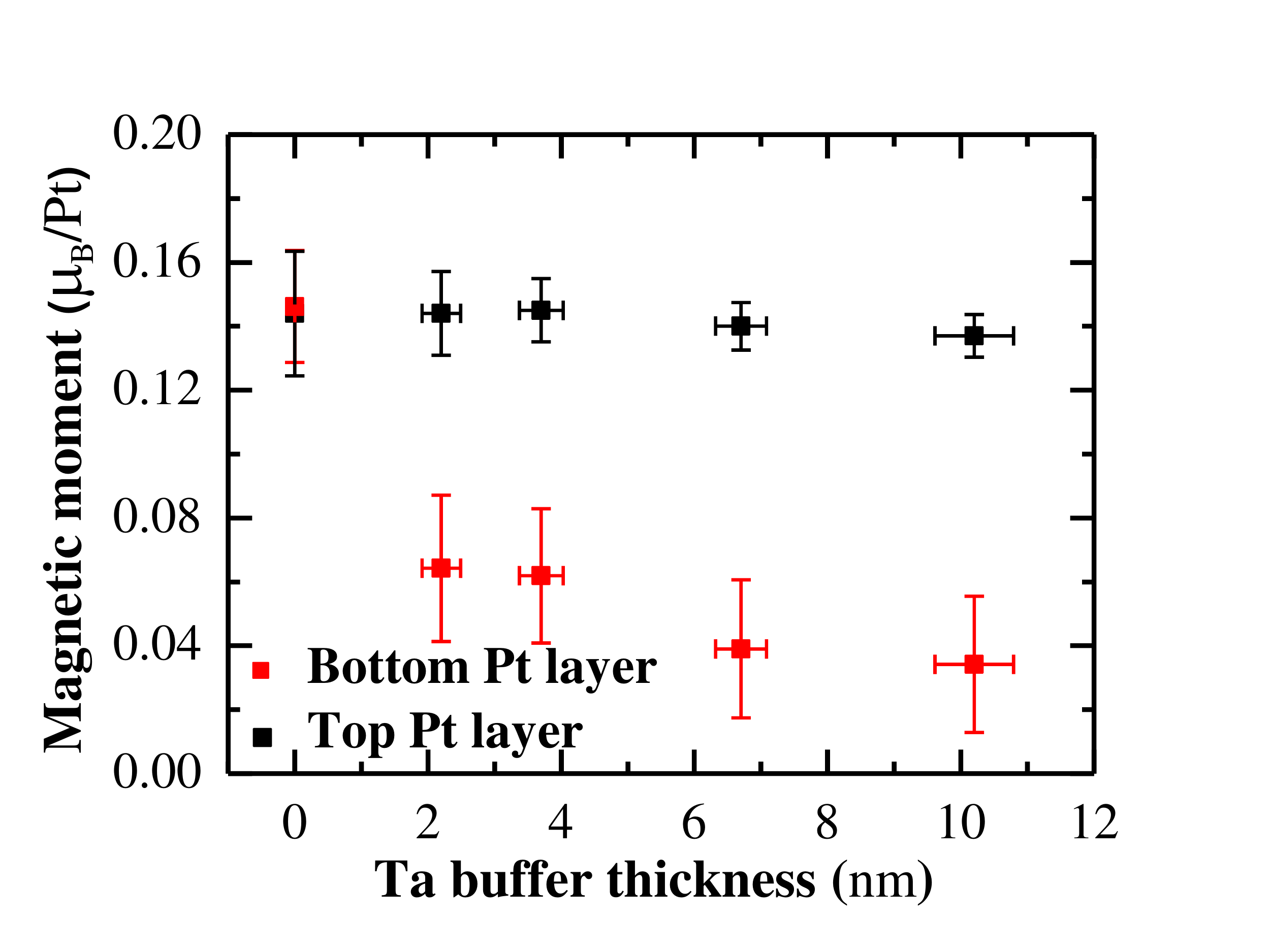}
\vspace{-6mm}
        \caption{(Color online) Dependence of magnetic moment in top and
bottom Pt layers on the Ta buffer thickness.}
    \label{fig:5}
\end{figure}

\begin{figure*}[t]
\centering
    \includegraphics[scale=0.38]{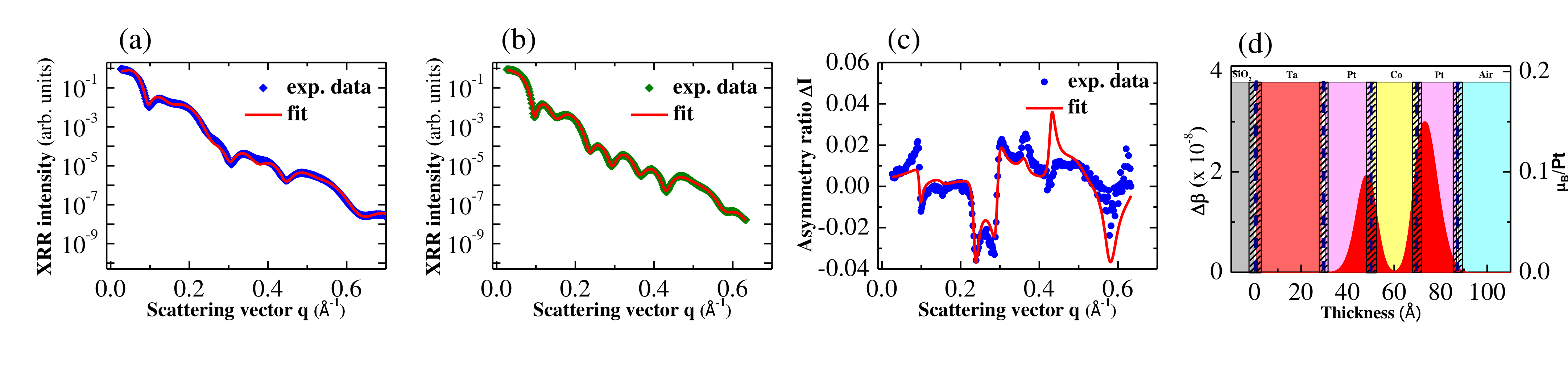}
    \vspace{-5mm}
        \centering
    \caption{(Color Online) (a) The off-resonant (11467 eV) averaged XRR signals with fits. (b) The resonant (11567
eV) averaged XRR signals with fits. (c) The corresponding measured and fitted XRMR asymmetry ratios $\Delta$I(q) for
all multilayers. (d) Magnetooptic depth profiles which were used to fit the XRMR asymmetry ratios. Red histograms
depict the in-depth distribution of the spin-polarized Pt atoms. Induced magnetic moment per Pt atom has been
quantified from \textit{ab initio} calculation. The dashed lines denote the corresponding interface position between the
layers whereas the dashed rectangles between the layers denote the corresponding interlayer roughness.}
    \label{fig:6}
\end{figure*}

\begin{figure}[htb!]
\centering
        \includegraphics[scale=0.3]{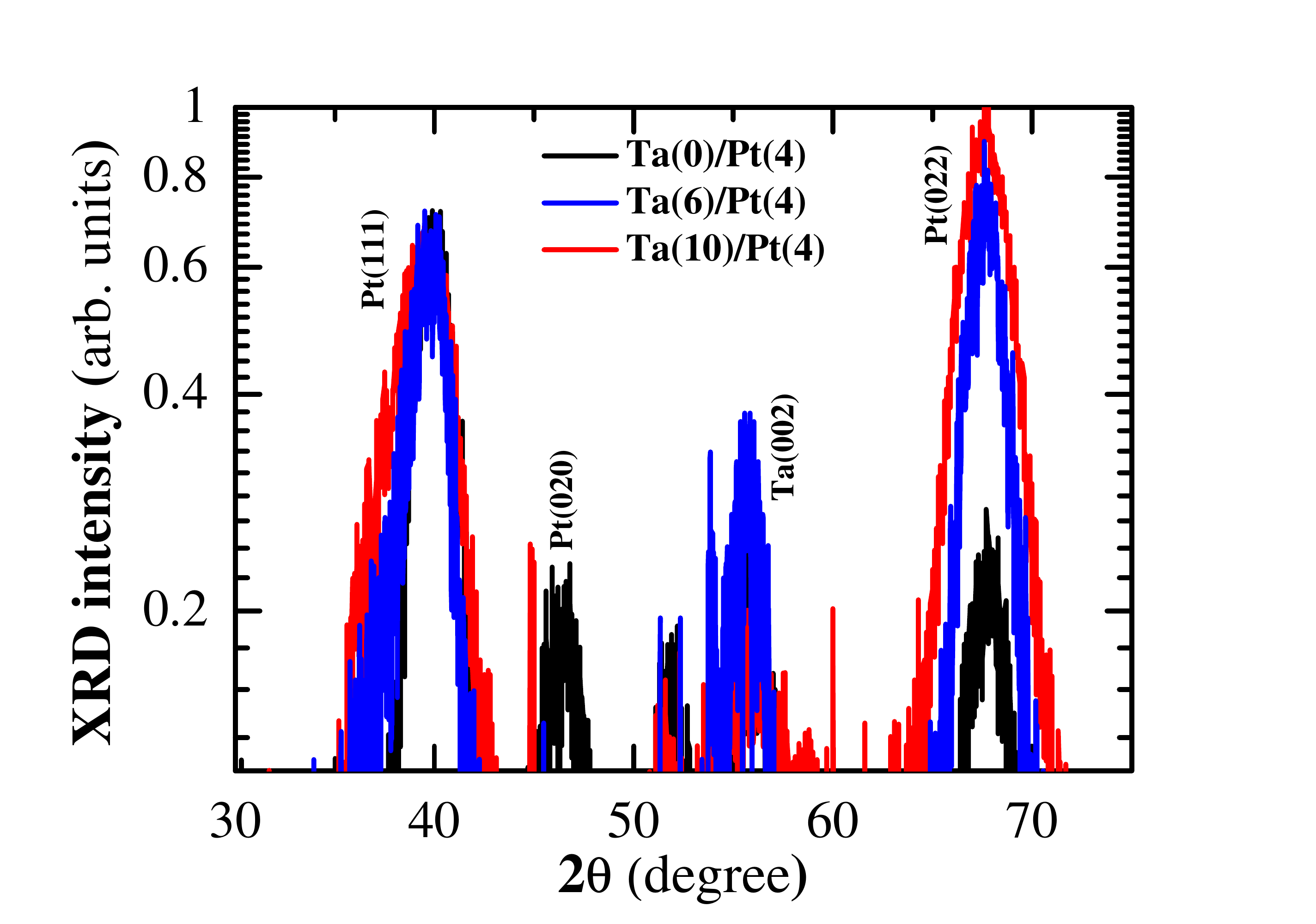}
        \caption{(Color online) Grazing incidence x-ray diffraction data with an incidence angle of 0.5\textsuperscript{o} of
Ta(\textit{t})/Pt(4nm) bilayers with \textit{t} = 0, 6, 10 nm.}
        \label{fig:7} 
\end{figure}

\begin{figure}[htb!]
        \includegraphics[trim={3.5cm 0 0 0},clip,width=0.55\textwidth] {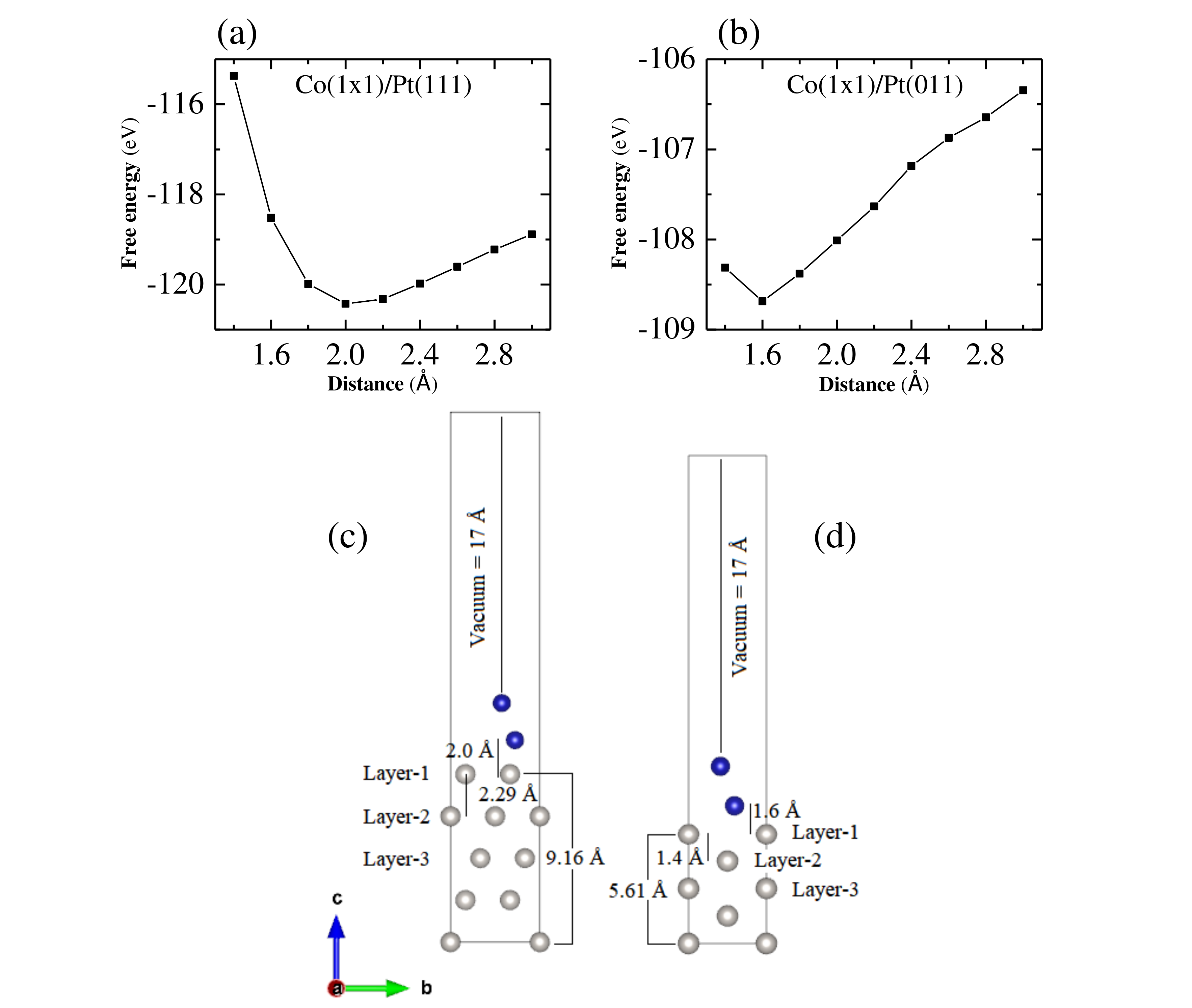}
        \centering
    \caption{(Color online) The free energy variation with the separation between Co and Pt layers in (a)
Co(1$\times$1)/Pt(111) and (b) Co(1$\times$1)/Pt(011), respectively. The modelled bilayers of Co/Pt as (c) Co (1$\times$1) on
Pt(111) and (d) Co (1$\times$1) on Pt(011). Grey and blue atoms are Pt and Co, respectively.}
    \label{fig:8}
\end{figure}

\begin{figure}[htb!]
\vspace{-2mm}
\centering

\includegraphics[scale=0.25]{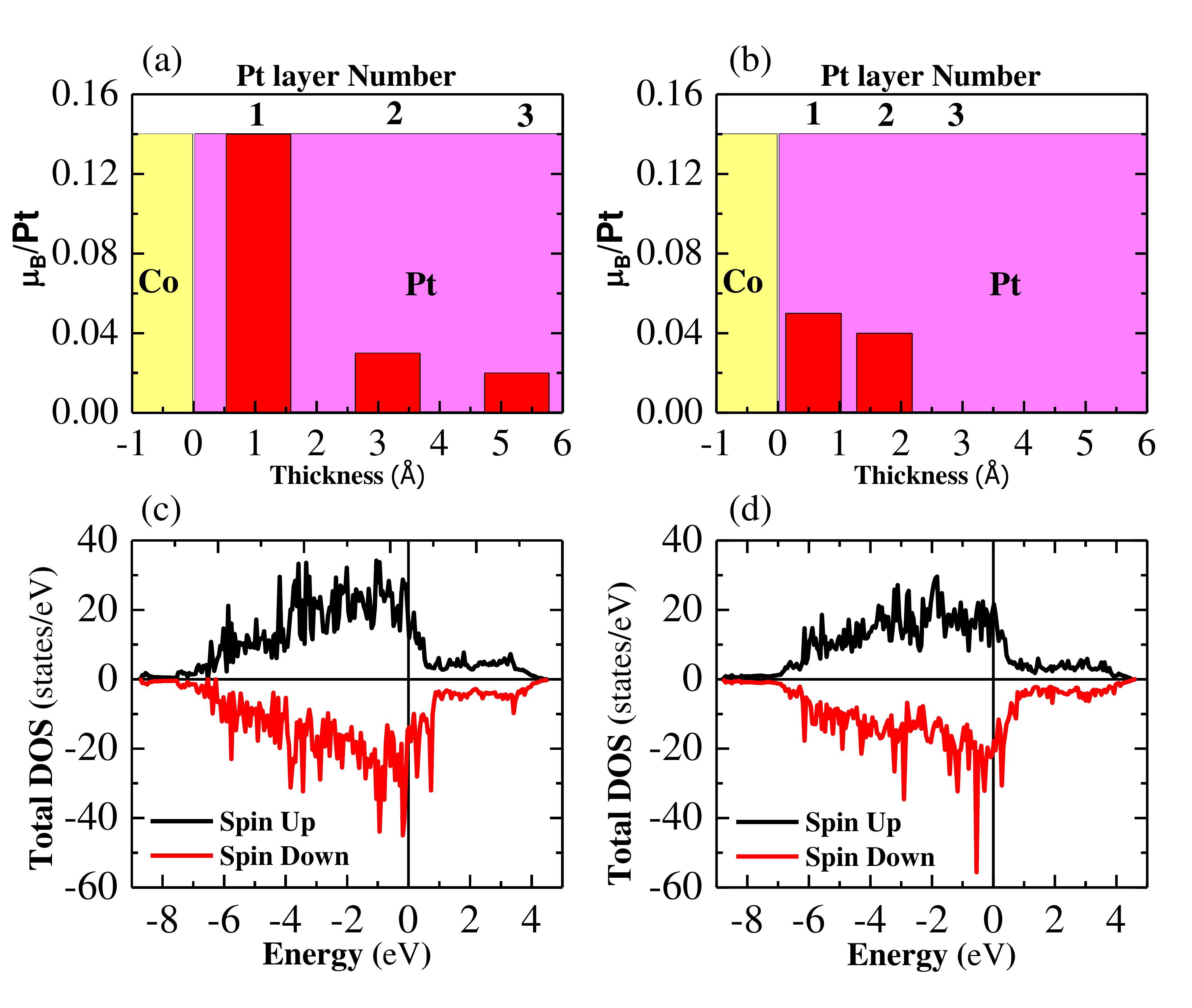}

        \caption{(Color online) Layer wise variation of the induced magnetic
moment (a) Co/Pt(111) and (b) Co/Pt(011) systems are
presented. The total density of states of (c) Co/Pt(111) and (d)
Co/Pt(011) systems are shown.}
    \label{fig:9}
\end{figure}
\noindent \\ Comparing the FWHMs of the spatial distribution of the induced magnetic moment due to the MPE it can be realized that the effective spin-polarized Pt thickness is about 1 nm for all the multilayers (see Tab. II) and it is independent of the Pt thickness confirming earlier results \cite{r16}. Comparing all the magnetic depth profiles of the spin-polarized Pt layers it can be realized that the maximum induced magnetic moment per Pt atom in the bottom Pt layer decreases with the increase in thickness of the Ta buffer layer (see Fig. 5). Moreover, XRMR has been also carried out on Ta(3)/Pt(2)/Co(2)/Pt(2) multilayer (see Fig. 6(c)). Due to the Ta buffer layer, the induced magnetic moments in the top and the bottom Pt layers are different, even if the thickness of the two Pt layers are the same (see Fig. 6(d)). It should be noted that the nearly equal roughness at the top Co/Pt and bottom Pt/Co interfaces rules out the possibility of different intermixing at the top and bottom Co/Pt interfaces to get different MPE in the two Pt layers (see Tab. I)). Therefore, the only possibility of this different moment value could be the difference in structural properties of the two Pt layers. Obviously, the structural properties of bottom Pt layer deposited on the Ta buffer layer could be different than that of top Pt layer deposited on the Co layer due to different growth ambience \cite{r52}.\\

\noindent
It is interesting to note that in recent years, hard XRMR studies on trilayer systems have been reported for Pt/Co/Pt \cite{r23} and Pd/Co/Pd \cite{r53}. It has been shown qualitatively that the top NM layer has an enhanced induced magnetic moment compared to bottom one. Hence, it is important to understand the origin of such an observation. Therefore, we deposited Ta(t)/Pt(4) bilayers on the same substrate under the same growth condition. Grazing incidence x-ray diffraction has been carried out on these bilayers with an incidence angle of 0.5\textsuperscript{o} (see Fig. 7) to get more structural information of the bottom Pt layer. It has been observed that with larger Ta thickness the growth of Pt(011) becomes more prominent than Pt(111).\\
\noindent We have done density functional theory calculations to compare between Co/Pt(111) and Co/Pt(011) interfaces. As the Ta buffer layer only modifies the crystallinity of the bottom Pt layer the Co/Pt interface is modelled accordingly. Here, Pt(111) and Pt(011) has been taken as substrates and ionic relaxation was performed for both the Pt substrates. A Co supercell of $1\times1$ was placed on top of both Pt(111) and Pt(011) substrates and vacuum of 17\si{\angstrom} was provided along the z-direction. The modelled Co/Pt(111) and Co/Pt(011) bilayers with optimized separation between Co and Pt layers are shown (see Fig. 8). In the modelled bilayers, only Co atoms were allowed to relax in the x-y plane with optimized z. We have used DFT as employed in Vienna Ab initio Simulation Package \cite{r54,r55,r56} to study the electronic structure and magnetic properties of the Co/Pt bilayers. Furthermore, a Perdew-Burke-Ernzerhof-based \cite{r57} spin-polarized semi-local exchange-correlation functional was used. Projected augmented wave \cite{r58,r59} method-based potentials has been used for both Co and Pt atoms. The valence electrons taken for the Co atoms and the Pt atoms were \textit{d\textsuperscript{8}s\textsuperscript{1}} and \textit{s\textsuperscript{1}d\textsuperscript{9}}, respectively. For structural optimization, Gaussian smearing was used with Monkhorst-Pack (MP) \cite{r60} \textbf{k}-mesh of $3\times5\times1$. The accurate calculation of electronic structure was carried out with tetrahedron smearing using finer MP \textbf{k}-mesh of $5\times11\times1$ and $5\times13\times1$ for Co/Pt(111) and Co/Pt(011) bilayer systems, respectively. The convergence criteria for forces on each atom in the structural optimization and for the electronic self-consistency was set to 0.0005 eV/\si{\angstrom} and $10^{-7}$ eV, respectively. The variation of the induced magnetic moment due to the Co atoms on the Pt atoms for both crystal structures in various layers are presented in Figs. 9(a) and 9(b) for Co/Pt(111) and Co/Pt(011), respectively. It can be seen that the induced magnetic moments are 0.14 $\mu$\textsubscript{B}/Pt and 0.05 $\mu$\textsubscript{B}/Pt at the interfaces of Co/Pt(111) and Co/Pt(011), respectively.\\
\noindent
Any difference in the structural or magnetic properties at the interface can impact the MPE to modify the spin-transport phenomena \cite{r20,r21} in NM/FM/NM systems. As the MPE originates from the 3\textit{d}-4\textit{d} or 3\textit{d}-5\textit{d} \cite{r32} hybridization, it can significantly affect the spin-transport properties in NM-FM systems, which is sensitive to the interfacial spin orbit coupling. Moreover, the asymmetry in the induced magnetic region of NM/FM and that of FM/NM can be manipulated to control the spin-dependent transport phenomena in spin-orbitronic devices \cite{r61}. \\
\section*{\centering\normalsize{{IV. CONCLUSION}}}
\noindent In conclusion, we investigated the magnetic proximity effect due to Co layer in two Pt layers of Ta/Pt/Co/Pt multilayers using XRMR. Analyzing the XRMR asymmetry ratio, it has been realized that the value of induced magnetic moments in top and bottom Pt layers are different irrespective of their structural thickness and interfacial roughness. The difference in the induced magnetic moment can give rise to various spin-transport phenomena such as SOT, iDMI. The Ta buffer layer promotes the growth of Pt(011) rather than Pt(111) as its thickness increases. Density functional theory calculations show that indeed, the Pt(111) layer has more induced magnetic moment than that of the Pt(011) due to the proximity of the Co layer. Moreover, it has been established that the magnetic moment of a multilayer is the summation of the moment in the FM layer and the moment induced by the FM layer in the adjacent NM layers.\\
\newline Major part of this work was carried out at the light source PETRA III, DESY, a member of the Helmholtz Association (HGF). We would like to thank David Reuther and Philipp Glaevecke, beamline engineer at the P09 beamline, PETRA III at DESY, for the technical support.  Financial support by the Department of Science and Technology (Government of India) provided within the framework of the India@DESY collaboration and Deutsche Forschungsgemeinschaft (DFG) within the individual grant RE 1052/42-1 are gratefully acknowledged. AM and SKV thank Ministry of Human Resource Development, Govt. of India and INSPIRE faculty scheme, DST, Govt. of India, respectively for the financial support. 

\newpage
%


\begin{thebibliography}{99}
%
\bibitem{r1} 
M. I. Dyakonov, and V. I. Perel, Phys. Lett. A \textbf{35}, 459 (1971).
\bibitem{r2}	S. Zhang, Phys. Rev. Lett. \textbf{85}, 393 (2000).
\bibitem{r3}	
L. Wang, R. J. H. Wesselink, Y. Liu, Z. Yuan, K. Xia, and P. J. Kelly, Phys. Rev. Lett. \textbf{116}, 196602 (2016).
\bibitem{r5}	J. Sinova, S. O. Valenzuela, J. Wunderlich, C. H. Back and T. Jungwirth, Rev. Mod. Phys. \textbf{87}, 1213 (2015).
\bibitem{r8}	I. E. Dzyaloshinskii, Sov. Phys. JETP \textbf{5}, 1259 (1957).
\bibitem{r9}	T. Moriya, Phys. Rev. \textbf{120}, 91 (1960).
\bibitem{r10}	R. Lavrijsen, D. M. F. Hartmann, A. van den Brink, Y. Yin, B. Barcones, R. A. Duine, M. A. Verheijen, H. J. M. Swagten, and B. Koopmans, Phys. Rev. B \textbf{91}, 104414 (2015).
\bibitem{r11}	I. M. Miron, T. Moore, H. Szambolics, L. D. Buda-Prejbeanu, S. Auffret, B. Rodmacq, S. Pizzini, J. Vogel, M. Bonfim, A. Schuhl and G. Gaudin, Nat. Mater. \textbf{10}, 419 (2011).
\bibitem{r12}	A. Manchon, H. C. Koo, J. Nitta, S. M. Frolov and R. A. Duine, Nat. Mater. \textbf{14}, 871 (2015).
\bibitem{r13}	P. Gambardella and I. M. Miron, Phil. Trans. R. Soc. A \textbf{369}, 3175 (2011).
\bibitem{r14}	I. M. Miron, K. Garello, G. Gaudin, P.-J. Zermatten, M. V. Costache, S. Auffret, S. Bandiera, B. Rodmacq, A. Schuhl and P. Gambardella, Nature \textbf{476}, 189 (2011).
\bibitem{r15}	J. Vogel,A. Fontaine, V. Cros, F. Petroff, J.-P. Kappler, G. Krill, A. Rogalev, and J. Goulon, Phys. Rev. B \textbf{55}, 3663 (1997).
\bibitem{r16}	T. Kuschel, C. Klewe, J.-M. Schmalhorst, F. Bertram, O. Kuschel, T. Schemme, J. Wollschläger, S. Francoual, J. Strempfer, A. Gupta, M. Meinert, G. Götz, D. Meier, and G. Reiss, Phys. Rev. Lett. \textbf{115}, 097401 (2015). 
\bibitem{r17}	P. K. Manna and S. M. Yusuf, Phys. Rep. \textbf{535}, 61 (2014).
\bibitem{r18}	M. J. Zuckermann, Solid State Commun. \textbf{12}, 745 (1973).
\bibitem{r19}	E. C. Stoner, Proc. R. Soc. \textbf{165}, 372 (1938).
\bibitem{r20}	P. M. Haney, H.-W. Lee, K.-J. Lee, A. Manchon, and M. D. Stiles, Phys. Rev. B \textbf{88}, 214417 (2013).
\bibitem{r21}	M. Jamali, K. Narayanapillai, X. Qiu, L. M. Loong, A. Manchon, and H. Yang, Phys. Rev. Lett. \textbf{111}, 246602 (2013).
\bibitem{r22}	W. Zhang, M. B. Jungfleisch, W. Jiang, Y. Liu, J. E. Pearson, S. G. E. te Velthuis, and A. Hoffmann, Phys. Rev. B \textbf{91}, 115316 (2015).
\bibitem{r23}	R. M. Rowan-Robinson, A. A. Stashkevich, Y. Roussigné, M. Belmeguenai, S.-M. Chérif, A. Thiaville, T. P. A. Hase, A. T. Hindmarch and D. Atkinson, Sci. Rep. \textbf{7}, 16835 (2017).
\bibitem{r24}	T. A. Peterson, A. P. McFadden, C. J. Palmstrøm, and P. A. Crowell, Phys. Rev. B \textbf{97}, 020403(R) (2018).
\bibitem{r25}	L. J. Zhu, D. C. Ralph and R. A. Buhrman, Phys. Rev. B \textbf{98}, 134406 (2018).
\bibitem{r26}	H. Maruyama, A. Koizumi, K. Kobayashi, and H. Yamazaki, Jpn. J. Appl. Phys. \textbf{32}, 290 (1993).
\bibitem{r27}	G. Schütz, R. Wienke, W. Wilhelm, W. B. Zeper, H. Ebert, and K. Spörl, J. Appl. Phys. \textbf{67}, 4456 (1990).
\bibitem{r28}	G. Schütz, R. Wienke, W. Wilhelm, W. Wagner, R. Frahm, and P. Kienle, Physica B \textbf{158}, 284 (1989).
\bibitem{r29}	W. J. Antel, Jr., M. M. Schwickert, T. Lin, W. L. O'Brien and G. R. Harp, Phys. Rev. B \textbf{60}, 12933 (1999).
\bibitem{r30}	F. Wilhelm, P. Poulopoulos, G. Ceballos, H. Wende, K. Baberschke, P. Srivastava, D. Benea, H. Ebert, M. Angelakeris, N. K. Flevaris, D. Niarchos, A. Rogalev, and N. B. Brookes, Phys. Rev. Lett. \textbf{85}, 413 (2000).
\bibitem{r31}	P. Poulopoulos, F. Wilhelm, H. Wende, G. Ceballos, K. Baberschke, D. Benea, H. Ebert, M. Angelakeris, N. K. Flevaris, A. Rogalev, and N. B. Brookes, J. Appl. Phys. \textbf{89}, 3874 (2001).
\bibitem{r32}	F. Wilhelm, P. Poulopoulos, A. Scherz, H. Wende, K. Baberschke, M. Angelakeris, N. K. Flevaris, J. Goulon, and A. Rogalev, Phys. Status Solidi A \textbf{196}, 33 (2003).
\bibitem{r33}	M. Suzuki, H. Muraoka, Y. Inaba, H. Miyagawa, N. Kawamura, T. Shimatsu, H. Maruyama, N. Ishimatsu, Y. Isohama, and Y. Sonobe, Phys. Rev. B \textbf{72}, 054430 (2005).
\bibitem{r34}	S. Rüegg, G. Schütz, P. Fischer, R. Wienke, W. B. Zeper, and H. Ebert, J. Appl. Phys. \textbf{69}, 5655 (1991).
\bibitem{r40}	C. Klewe, T. Kuschel, J.-M. Schmalhorst, F. Bertram, O. Kuschel, J. Wollschläger, J. Strempfer, M. Meinert, and G. Reiss Phys. Rev. B \textbf{93}, 214440 (2016).
\bibitem{r41}	S. Macke, and E. Goering, J. Phys.: Condens. Matter \textbf{26}, 363201 (2014).
\bibitem{Sonia}	J. Strempfer, S. Francoual, D. Reuther, D. K. Shukla, A. Skaugen, H. Schulte-Schrepping, T. Kracht and H. Franz, J. Synchrotron Radiat. \textbf{20}, 541 (2013).
\bibitem{r43}	S. Macke, ReMagX, An x-ray magnetic reflectivity tool, \footnotesize{www.remagx.org}.
\normalsize
\bibitem{r44}	L. G. Parratt, Phys. Rev. \textbf{95}, 359 (1954).
\bibitem{r45}	L. Névot and P. Croce, Rev. Phys. Appl. \textbf{15}, 761 (1980).
\bibitem{r46}	J. Zak, E. R. Moog, C. Liu, and S. D. Bader, J. Magn. Magn. Mater. \textbf{89}, 107 (1990).
\bibitem{r47}	T. Kuschel, C. Klewe, P. Bougiatioti, O. Kuschel, J. Wollschläger, L. Bouchenoire, S. D. Brown, J.-M. Schmalhorst, D. Meier, G. Reiss, IEEE Trans. Magn. \textbf{52}, 4500104 (2016).
\bibitem{r48}	P. Bougiatioti, O. Manos, O. Kuschel, J. Wollschläger, M. Tolkiehn, S. Francoual, T. Kuschel, arXiv:1807.09032 (2018).
\bibitem{r49}	A. V. Kolobov, F. Wilhelm, A. Rogalev, T. Shima, and J. Tominaga, Appl. Phys. Lett. \textbf{86}, 121909 (2005).
\bibitem{r51}	R. I. Allen, and F. W. Constant, Phys. Rev. \textbf{44}, 228 (1933).
\bibitem{r52}	R. Lavrijsen, P. P. J. Haazen, E. Mure, J. H. Franken, J. T. Kohlhepp, H. J. M. Swagten, and B. Koopmans, Appl. Phys. Lett. \textbf{100}, 262408 (2012).
\bibitem{r53}	D.-O. Kim, K. M. Song, Y. Choi, B.-C. Min, J.-S. Kim, J. W. Choi, and D. R. Lee, Sci. Rep. \textbf{6}, 25391 (2016).
\bibitem{r54}	G. Kresse and J. Furthmüller, Comput. Mater. Sci. \textbf{6}, 15 (1996).
\bibitem{r55}	G. Kresse and J. Furthmüller, Phys. Rev. B \textbf{54}, 11169 (1996).
\bibitem{r56}	G. Kresse and D. Joubert, Phys. Rev. B \textbf{59}, 1758 (1999).
\bibitem{r57}	J. P. Perdew, K. Burke, and M. Ernzerhof, Phys. Rev. Lett. \textbf{77}, 3865 (1996).
\bibitem{r58}	P. E. Blöchl, Phys. Rev. B \textbf{50}, 17953 (1994).
\bibitem{r59}	G. Kresse and D. Joubert, Phys. Rev. B \textbf{59}, 1758 (1999).
\bibitem{r60}	H. J. Monkhorst and J. D. Pack, Phys. Rev. B \textbf{13}, 5188 (1976).
\bibitem{r61}	R. Ramaswamy, J. M. Lee, K. Cai, and H. Yang, arXiv:1808.06829 (2018).

%

%

%
\end{thebibliography}
\end{document}